 \documentclass{iopart}            %%%% COMENTAR PARA ENVIAR
\usepackage{graphicx}% Include figure files
\usepackage{harvard}
  \usepackage{iopams}

\begin{document}

\title[The Winfree model with heterogeneous phase-response curves]
{The Winfree model with heterogeneous phase-response curves: Analytical results}

\author{Diego Paz\'o$^1$, Ernest Montbri\'o$^2$ and Rafael Gallego$^3$}
\address{$^1$ Instituto de F\'isica de Cantabria (IFCA), CSIC-Universidad de Cantabria, 39005 Santander, Spain}
\address{$^2$ Department of Information and Communication Technologies,
Universitat Pompeu Fabra, 08018 Barcelona, Spain}
\address{$^3$Departamento de Matem\'aticas, Universidad de Oviedo, Campus de
Viesques, 33203 Gij\'on, Spain}

\begin{abstract}
We study an extension of the Winfree model of coupled phase oscillators in
which both natural frequencies and phase-response curves
(PRCs) are heterogeneous. 
In the first part of the paper we resort to averaging and 
derive  an approximate model, 
in which the oscillators are coupled through
their phase differences. Remarkably, this simplified model is the `Kuramoto
model with distributed shear' (2011 {\sl Phys.~Rev.~Lett.}~{\bf 106} 254101). 
We find that above a critical level of PRC heterogeneity the incoherent state is always stable. 
In the second part of the paper we perform the analysis of the full model for
Lorentzian heterogeneities, resorting to the Ott-Antonsen ansatz. 
The critical level of PRC heterogeneity obtained within the averaging approximation
has a different manifestation in the full model 
depending on the sign of the center of the distribution of PRCs.
\end{abstract}

\vspace{2pc}

% \submitto{\jpa}
% Comment out if separate title page not required  %%%%%%%%%%%%%%%%%%%%%
% \maketitle

\section{Introduction}

The Winfree model \cite{Win67,Win80} is a milestone in the mathematical description of 
collective synchronization.  Inspired  by the 
synchronization of biological oscillators, 
Winfree proposed a model  
consisting of a large population ($N\gg1$) of interacting limit cycle oscillators with 
heterogeneous natural frequencies, 
capable of self-synchronizing macroscopically, see e.g.~\cite{Str03}. As simplifying assumptions,
he prescribed that the phases $\theta_i$ ($i=1,\ldots,N$) 
were the only degrees of freedom, and that the interactions
were equally weighted and global,
i.e.~mean-field type. Despite its deep conceptual influence, the theoretical description of the Winfree model 
remains a challenging problem and analytical progress is scarce, see e.g.~\cite{AS01,quinn,PM14,PR15,Gallego2017}.

In the Winfree model, each oscillator 
responds to the incoming pulses according to the value 
of the phase-response curve (PRC). Specifically, the 
PRC ---also called infinitesimal PRC, phase resetting curve, or sensitivity function \cite{Izh07}--- is a function of only the oscillator's own phase, 
and determines the advance or delay of its phase in response to a certain 
perturbation.
The PRC plays a fundamental role in neuroscience \cite{smeal10,prcbook}, 
and it has been determined experimentally
in cortical neurons \cite{reyes93,reyes93b,netoff05,tateno07,tsubo_ejn07,mancilla07}, hippocampal neurons  
\cite{lengyel05},
mitral cells \cite{galan05}, or in neurons of the abdominal 
ganglia of Aplysia \cite{preyer05}.
Additionally, synchronization 
of biological oscillators 
such as fireflies \cite{buck88}, tropical katydids \cite{sismondo90} and 
the human heart \cite{kralemann} have been analyzed through PRCs. 
The concept of PRC is also important 
for technological applications such as electric oscillators \cite{HL98}
or wireless sensor nets, see e.g.~\cite{nishimura11} and references therein. 

In its original form the Winfree model is made of oscillators with heterogeneous 
natural frequencies.
Yet, it is reasonable to assume that heterogeneity may well also 
be present in other system's parameters, and that this may influence synchronization. 
Indeed, broad cell-to-cell differences 
in PRCs have been recently measured in the olfactory bulb mitral cells~\cite{burton12},
and given that the collective phase dynamics of a synchronized 
ensemble of oscillators depends crucially on the level of PRC heterogeneity
~\cite{nakao18}, it is desirable to deepen 
our understanding on the effects of heterogeneous PRCs on
collective synchronization. However, due to its mathematical complexity,
previous attempts to tackle oscillator ensembles with heterogeneous PRCs are scarce, and 
rely on approximate methods \cite{tsubo07,ly14}. 

In this paper, we study an extension of the classical Winfree model with heterogeneous natural frequencies and PRCs.
In section \ref{sec:model} we present the model. An approximate version of it, 
based on averaging, is analyzed in section \ref{sec:av}.
Section \ref{sec:oa} presents exact results obtained by means of the Ott-Antonsen theory. Finally, in section \ref{sec:concl} we summarize
the main conclusions of our work, and suggest future lines of research.

\section{Model description}
\label{sec:model}

The Winfree model consists of an ensemble of $N\gg1$ all-to-all coupled
phase oscillators whose phases $\theta_i$ ($i=1,\dots,N$)
evolve according to the following set of $N$ coupled ordinary differential
equations (ODEs):
\begin{equation}\label{eq1}
  \dot\theta
  _i=\omega_i+Q_i(\theta_i)\frac{\varepsilon}{N}\sum_{j=1}^NP(\theta_j).
\end{equation}
Here $\omega_i$ are the natural frequencies, and 
$\varepsilon>0$ is a parameter controlling the coupling strength. 
The function $P$ specifies the form of the pulses, and
the response of the $i$-th
oscillator to the mean field $N^{-1} \sum_j P(\theta_j)$ is determined by the PRC function $Q_i(\theta)$. 

\begin{figure}
  \includegraphics*[width=120mm]{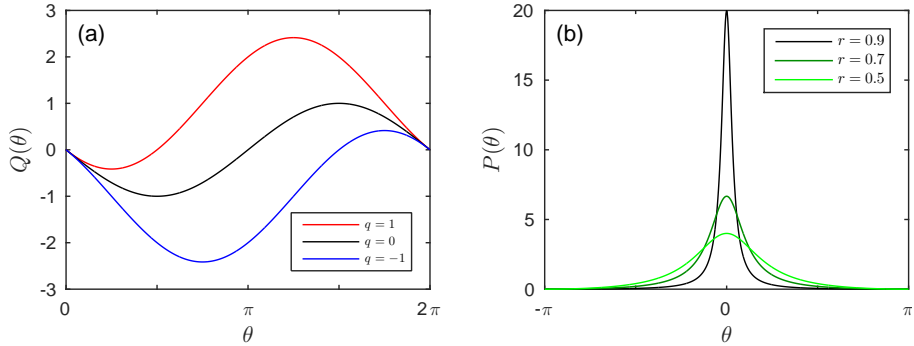}
  \caption{ (a) Phase response curve and (b) pulse shape for three representative values of $q$ and $r$, respectively.}
  \label{fig::qp}
\end{figure}

Note that in (\ref{eq1}), the subscript $i$ appears twice in the right hand-side:
in the natural frequencies, and in the PRCs. 
Specifically, we consider the monoparametric family of PRCs
\begin{equation}
  Q_i(\theta)=q_i(1-\cos\theta)-\sin\theta,
  \label{prc}
\end{equation}
where parameter $q_i$ controls if the PRC is more positive than negative ($q_i>0$),
or the other way around ($q_i<0$), see figure \ref{fig::qp}(a). 
With the parametrization adopted here we have $Q_i(0)=0$, 
since we choose $\theta=0$ as the point where the pulse peaks.

The pulse $P(\theta)$ is assumed to be a symmetric unimodal function in the interval $[-\pi,\pi]$,
with the normalization $\int_{-\pi}^{\pi} P(\theta) \rmd\theta=2\pi$.
In section \ref{sec:oa}, 
we adopt the pulse function \cite{Gallego2017}:
\begin{equation}
P(\theta)=\frac{(1-r)(1+\cos\theta)}{1-2r\cos\theta +r^2},
\label{p}
\end{equation}
which vanishes at $\theta=\pi$.
Parameter $r$, controlling the width of the pulse, spans between
$-1$ (flat pulse) and $1$ (Dirac-delta pulse, $P(\theta)=2\pi\delta(\theta)$), 
see examples in figure \ref{fig::qp}(b).

In section \ref{sec:av} we study the approximation of \eref{eq1}-\eref{prc} based on averaging. The results
in that section depend exclusively on the first Fourier mode of $P(\theta)$, and not on the other features of the pulse. 
The specific pulse type is, nonetheless, relevant for the exact results in section \ref{sec:oa}.
Our study is focused on determining the parameter values where the completely asynchronous state is unstable,
making certain level of synchrony unavoidable. By synchrony, we refer to a state in which a macroscopic fraction of
the ensemble is entrained to the same frequency and remains phase locked.

%==========================
\section{Averaging approximation}
\label{sec:av}
%==========================

In this section we analyze an approximation of the Winfree model with heterogeneous PRCs, which is particularly  amenable to theoretical analysis.
This permits to study general distributions of $\omega$ and $q$, and at the same time, the results obtained
serve as a guide for section \ref{sec:oa}, where an exact analysis is presented.
Using the method of averaging \cite{Kur84}, valid
for weak coupling and small frequency dispersion,
the system of $N$ ODEs \eref{eq1} may be simplified to a model where interactions 
are described exclusively by phase differences. For the PRCs in (\ref{prc}) we find: 
\begin{equation}
\dot \theta_i=\omega_i +\varepsilon q_i
+ \Pi  \frac{\varepsilon}{N}
\sum_{j=1}^N [\sin(\theta_j-\theta_i)
-q_i \cos(\theta_j-\theta_i)], 
\label{av_winfree}
\end{equation}
at the lowest order in $\varepsilon$.
The sinusoidal shape of the PRCs is responsible of (i) the absence of
higher harmonics in the coupling functions, and (ii) the presence of the constant
$\Pi=\hat p_1$, a `shape factor' that equals the first Fourier mode of the pulse $P(\theta)=\sum_{n=-\infty}^\infty \hat p_n \rme^{\rmi n\theta}$.
Specifically, for the pulse type in \eref{p}, 
\begin{equation}
\Pi=\frac{1+r}{2}, 
\end{equation}
and therefore $0<\Pi<1$ for this and other pulses \cite{Gallego2017}.
The largest $\Pi$ value is $1$ and it is attained in the limit case of a Dirac delta Pulse, $P(\theta)=2\pi\delta(\theta)$. 
Remarkably, in this limit the model in \eref{av_winfree} coincides with the `Kuramoto model 
with distributed shear', which was originally 
deduced as a phase approximation for globally coupled
Stuart-Landau oscillators with distributed natural frequencies and shears 
(or nonisochronicities) \cite{MP11}. 
Instead, here the Kuramoto model  
\eref{av_winfree} is obtained 
from the Winfree model with heterogeneous PRCs.
This coincidence permits to transfer the results from \cite{MP11} for $\Pi=1$, 
or simply borrow the analysis used there for $\Pi<1$. 

In terms of the Kuramoto order parameter, $Z\equiv R\, \rme^{\rmi\psi}=N^{-1} \sum_j \rme^{\rmi\theta_j}$, model \eref{av_winfree}
can be alternatively written as,
\begin{equation}
\dot \theta_i=\omega_i +\varepsilon \, q_i
+ \Pi \, \varepsilon \, R
[\sin(\psi-\theta_i)
-q_i \cos(\psi-\theta_i)],
\label{av_winfree2}
\end{equation}
emphasizing in this way the mean-field character of the model.

\subsection{Linear stability analysis of incoherence}

Hereafter we only consider the thermodynamic limit ($N\to\infty$) of the model.
Hence, we define a phase density $f(\theta|\omega,q,t)$ of oscillators with
frequency $\omega$ and PRC-parameter $q$ at time $t$.  The mean field $Z$ in
this continuous formulation becomes
\begin{equation}
Z(t)=\int_{-\infty}^{\infty}\int_{-\infty}^{\infty}  p(\omega,q)
\int_{-\pi}^{\pi} f(\theta|\omega,q,t) \rme^{\rmi\theta} ~ \rmd\theta ~
\rmd\omega ~ \rmd q ,
\label{z_cont}
\end{equation}
where $p(\omega,q)$ is the joint probability distribution of $\omega$ and $q$.
In the uniform incoherent state the oscillators are uniformly scattered
in the unit circle, 
or otherwise said, $f$ equals $(2\pi)^{-1}$, and therefore the mean field vanishes, $Z=0$.
In an arbitrary state, $f$ is constrained to obey the continuity equation \cite{SM91,MP11}:
\begin{equation}
\partial_t f= - \partial_\theta
\left( \left\{\omega+\varepsilon q +  \frac{\Pi \varepsilon}{2\rmi}\left[Z \rme^{-\rmi \theta}(1 - \rmi q) - 
{\rm c.c.}\right] \right\} f \right)
\label{cont_eq}
\end{equation}
(${\rm c.c.}$ denotes complex conjugate), because of the conservation of the number of oscillators.
Note that this is a nonlinear equation since $Z$ depends on $f$ through \eref{z_cont}.
For the analysis that follows we write the Fourier series of $f$:
\begin{equation}
f(\theta|\omega,q,t)=\frac{1}{2\pi}\sum_{l=-\infty}^\infty {\hat
f}_l(\omega,q,t) \rme^{\rmi l\theta},
\label{four}
\end{equation}
with ${\hat f}_0=1$, and ${\hat f}_l={\hat f}_{-l}^*$.
We can insert \eref{four} into (\ref{cont_eq}) obtaining an infinite set of integro-differential equations that govern the
evolution of ${\hat f}_l$ in terms of itself, ${\hat f}_{l\pm1}$, and the mean field $Z$:
\begin{equation}
\partial_t{\hat f}_l= -\rmi(\omega + \varepsilon q) {\hat f}_l - 
\frac{\Pi \, \varepsilon \,l}{2} \left[Z {\hat f}_{l+1}(1-\rmi q)-Z^* {\hat f}_{l-1}(1+\rmi q) \right] ,
\label{fl}
\end{equation}
(the asterisk denotes the complex conjugation). It is crucial to note that, according to \eref{z_cont},  $Z$ depends only on the
first Fourier mode of the density:
\begin{equation}
Z^*(t)=\int_{-\infty}^{\infty}\int_{-\infty}^{\infty}  p(\omega,q) {\hat f}_1
(\omega,q,t) ~ \rmd\omega ~ \rmd q \equiv \left< {\hat f}_1 \right>. 
\label{z1}
\end{equation}
(The bracket is used hereafter to denote the  average over $\omega$ and $q$.)
Then, it can be easily verified in \eref{fl} that infinitesimal deviations from uniform incoherence (${\hat f}_{l\ne0}=0$) are governed solely by 
the first Fourier mode:
\begin{equation}
 \partial_t \hat f_1 = -\rmi(\omega+\varepsilon q) \hat f_1 + \frac{\Pi \, \varepsilon}{2}(1+\rmi q) 
%  \iint_{-\infty}^{\infty} p(\omega',q') 
 \left< \hat f_1 \right>.
%  d\omega' dq'
 \label{f1}
\end{equation}
A succession of well-known steps permits to determine the linear stability of incoherence \cite{SM91,Str00}: 
(i) insert the ansatz corresponding to an exponential growth rate $\lambda$, $\hat f_1=b(\omega,q) \rme^{\lambda t}$, in \eref{f1}; 
(ii) isolate $b$ in the left hand-side;
(iii) multiply both sides of the equation by $p(\omega,q)$; and (iv) integrate over $\omega$ and $q$. These steps yield a self-consistent
condition:
\begin{equation}
\frac{2}{\Pi\, \varepsilon}=
% \iint_{-\infty}^{\infty} 
 \left< \frac{1+\rmi q}{\lambda+\rmi (\omega+\varepsilon q)} \right>.
% \, p(\omega,q) \, d\omega \, dq 
\label{selfc}
\end{equation}
We can split this equation into a system of two equations for the imaginary and real parts:
\begin{equation}
\eqalign{
0 = 
% \iint_{-\infty}^\infty  
 \left<\frac{q \lambda_R-(\omega+\lambda_I+q \varepsilon)}{\lambda_R^2+ (\omega+\lambda_I+q \varepsilon)^2} \right> ,
% p(\omega,q) \, d\omega\, dq
 \cr
\frac{2}{\Pi \, \varepsilon} = 
% \iint_{-\infty}^\infty 
\left<
\frac{\lambda_R+q (\omega+\lambda_I+q \varepsilon)}{\lambda_R^2+ (\omega+\lambda_I+q \varepsilon)^2} %p(\omega,q) \, d\omega\, dq
\right>, }
\label{lsa}
\end{equation}
where $\lambda_R=\mathrm{Re}\,\lambda$  and $\lambda_I=\mathrm{Im}\, \lambda$.
For simplicity, we consider hereafter $ \omega$ and $q$ to be independently distributed, i.e.~$p(\omega,q)= g(\omega) h(q)$
---for correlated distributions with $\Pi=1$, see \cite{PM11}.
Moreover, it is convenient to assume that $g$ and $h$ are unimodal symmetric
functions \cite{SM91}. We can freely choose $g(\omega)$ centered at zero, since this can always be achieved by 
going to a rotating framework if necessary, while $h(q)$ is centered at a specific $q_0$ value. Note also that
changing the sign of $\omega$ and $q$ in \eref{selfc} transforms $\lambda$ into $\lambda^*$,
meaning that within the averaging approximation the sign of $q_0$ is irrelevant concerning the stability properties.
To compute the stability boundary we take the limit $\lambda_R\to 0^+$
in (\ref{lsa}), and obtain:
% \numparts
% \begin{eqnarray} 
\begin{equation}
\eqalign{0= 
% \iint_{-\infty}^\infty 
\left<\pi q  \delta(\lambda_I +q \varepsilon_c +\omega) - \frac{1}{\lambda_I+q \varepsilon_c+\omega} \right> \cr
% g(\omega) \, h(q)\, d\omega  \, dq \nonumber 
\frac{2}{\Pi\, \varepsilon_c}=
\left<\pi \delta(\lambda_I +q \varepsilon_c + \omega) + \frac{q}{\lambda_I+q\varepsilon_c+\omega} \right>}
\label{lsac}
\end{equation}
% \end{eqnarray}
% \endnumparts

\subsection{Lorentzian heterogeneities}
\begin{figure}
  \includegraphics*[width=120mm]{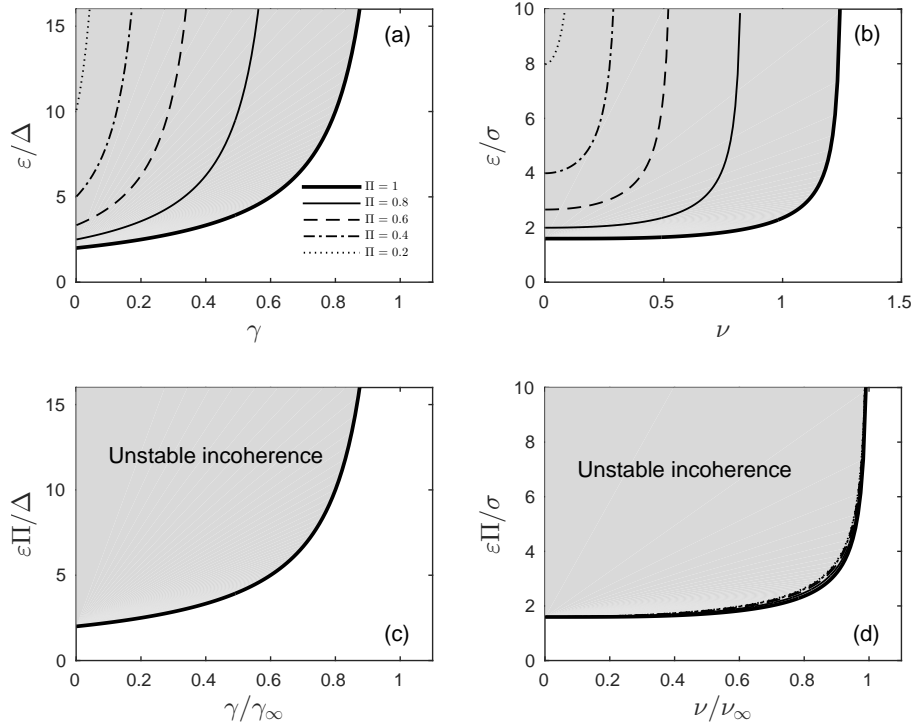}
  \caption{(a,b) Stability boundary of incoherence in the
  different values 
  of $\Pi$, with (a) Lorentzian, and (b) Gaussian PRC heterogeneities. In panel (a) the value of $q_0$ is irrelevant, while in panel 
(b) $q_0=0$.
The shaded regions correspond to unstable 
incoherence for $\Pi=1$. 
Panels (c,d) show the same boundaries 
after rescaling the axes. 
In the case of Gaussian heterogeneity, panel (d), the curves fail to collapse.}
  \label{fig::lg}
\end{figure}

For Lorentzian distributions 
\begin{equation}
g(\omega)= \frac{\Delta/\pi}{(\omega-\omega_0)^2+\Delta^2},~~h(q)=
\frac{\gamma/\pi}{(q-q_0)^2+\gamma^2},
\label{lorentzians}
\end{equation}
solving \eref{lsac} yields the critical coupling strength\footnote{In this case is perhaps easier to resort to the Ott-Antonsen ansatz, rather than
(\ref{lsac}). The result is obviously independent of the method chosen.} where incoherence becomes unstable
\begin{equation}
\varepsilon_c=  \frac{2\, \Delta}{\Pi-\gamma(2-\Pi)} ,
\label{av_c}
\end{equation}
which holds only for positive $\varepsilon$.  Notably, (\ref{av_c}) is independent of
$q_0$, a peculiarity of the Lorentzian distribution (in contrast to the independence on $\omega_0$ discussed above).
For a given $\Pi$ value, see figure \ref{fig::lg}(a), the function $\varepsilon_c(\gamma)/\Delta$ in (\ref{av_c}) 
defines a curve in the $(\gamma,\varepsilon/\Delta)$ plane that emanates from the $\varepsilon/\Delta$-axis 
at $2/\Pi$ and grows monotonically up
to a critical value
\begin{equation}
\gamma_\infty=\frac{\Pi}{2-\Pi},
\label{g_infty}
\end{equation}
where the curve diverges. In turn, incoherence is always stable 
for $\gamma>\gamma_\infty$.  As can be seen in figure \ref{fig::lg}(c), the formula \eref{av_c} can be condensed into a single curve
with rescaled variables:
\begin{equation}
\frac{\varepsilon_c \Pi}{\Delta}=\frac{2}{1-\gamma/\gamma_\infty}.
\label{resc}
\end{equation}

\subsection{Gaussian heterogeneities}

The analysis of distributions different from \eref{lorentzians} is more cumbersome. 
We consider here only Gaussian heterogeneities:
\begin{equation}
g(\omega)= \frac{1}{\sqrt{2\pi}\sigma} \rme^{-\omega^2/(2\sigma^2)},~~
h(q)=\frac{1}{\sqrt{2\pi}\nu} \rme^{-(q-q_0)^2/(2\nu^2)}.
\label{gaussians}
\end{equation}
The calculations are greatly simplified if 
$h(q)$ is centered at zero, i.e.~$q_0=0$.
In this case, after some manipulations of \eref{lsac}, we get a closed formula for the critical coupling
\begin{equation}
\varepsilon_c^{(q_0=0)} = \frac{4 \,\sigma}{\left[\sqrt{\pi \Pi^3
(16\nu^2+\pi\Pi)}+\pi \Pi^2 + 8\nu^2(\Pi-2)\right]^{1/2} }.
\label{ec2}
\end{equation}
For $\Pi=1$ we recover the result in \cite{MP11}. 
\Eref{ec2} defines a region in the $(\nu,\varepsilon/\sigma)$ plane that is maximal for $\Pi=1$ and progressively shrinks
as $\Pi$ is decreased, see figure \ref{fig::lg}(b) for several $\Pi$ values. 
The line (\ref{ec2}) is born at $\nu=0$ with $\varepsilon_c/\sigma=\sqrt{8/\pi}/\Pi$, 
and diverges at a critical value of $\nu$: 
\begin{equation}
\nu_\infty^{(q_0=0)}=\frac{\Pi}{2-\Pi}\sqrt{\frac{\pi}{2}}.
\label{nu_infty}
\end{equation}
\Eref{ec2} cannot be recast into a single formula valid for all $\Pi$ values, rescaling $\varepsilon_c$ and $\nu$. We see in
figure \ref{fig::lg}(d) that a rescaling analogous to \eref{resc} yields an imperfect collapse
of the boundaries.

Finally, we stress that our stability analysis is local, and hence
stable incoherence does not preclude its coexistence with a partially synchronized state, 
as it may occur for $\Pi=1$ see \cite{MP11}.

\subsection{Critical PRC heterogeneity}

The Lorentzian \eref{lorentzians} and  
Gaussian \eref{gaussians} joint distributions exhibit a critical
value of heterogeneity in $q$ such that, if $q$ is too heterogeneous, incoherence becomes stable for all $\varepsilon$.
Next, we investigate if a general rule ---for unimodal symmetric $h(q)$---  exists.
First of all we neglect the diversity of $\omega$ in (\ref{lsac}), since we are interested in the limit $\varepsilon_c\to\infty$. Intuitively, the term  $\varepsilon q_i $ in \eref{av_winfree}
can be as large in magnitude with respect to $\omega_i$ as desired. Mathematically, we can hence neglect the heterogeneity 
of $\omega$ taking $g(\omega)=\delta(\omega)$.
In addition, we rescale $\lambda_I$ by $\varepsilon_c$ and define $\lambda_I=\Lambda \varepsilon_c$.
In this way the dependence on $\varepsilon_c$ in \eref{lsac} cancels out, and we obtain 
the conditions:
\numparts
\begin{eqnarray}
0=-\pi \Lambda h_\infty(-\Lambda) - \int_{-\infty}^\infty  \frac{1}{\Lambda+q}
h_\infty(q) \, \rmd q,  \label{a}\\
2\, \Pi^{-1}= \pi h_\infty(-\Lambda) + \int_{-\infty}^\infty \frac{q}{\Lambda+q}
h_\infty(q) \, \rmd q. \label{b}
\end{eqnarray}
\endnumparts
Here, $h_\infty$ means the critical distribution of $h(q)$ such that 
the stability boundary is at $\varepsilon_c=\infty$. In other words, 
if $h(q)$ becomes infinitesimally broader, 
incoherence becomes stable for all $\varepsilon$. 
To get rid of the integral, we can multiply \eref{a} by $\Lambda$ and subtract \eref{b}
obtaining:
\begin{equation}
 \frac{2 \,  \Pi^{-1}-1}{\pi (1+\Lambda^2)}= h_\infty(-\Lambda). \label{primera}
\end{equation}
Additionally, multiplying \eref{b} by $\Lambda$ and adding \eref{a} yields after trivial manipulations:
\begin{equation}
\left(2 \, \Pi^{-1}-1\right)\Lambda=-(1+\Lambda^2) \int_{-\infty}^\infty
\frac{h_\infty(q-\Lambda)}{q}  \, \rmd q.
\label{segunda}
\end{equation}

\subsubsection{Centered $h(q)$ ($q_0=0$)} 

If $h(q)$ is centered at zero, symmetry imposes the trivial solution $\Lambda=0$
in \eref{segunda} (we are interpreting the integral in the Cauchy principal value sense). If $\Lambda>0$ the integral in \eref{segunda} is positive and the condition cannot be fulfilled,
likewise for $\Lambda<0$.
In consequence we get from \eref{primera} the remarkable result that 
the divergence of $\varepsilon_c$ is linked to a simple condition for the distribution maximum:
\begin{equation}
 h_\infty^{(q_0=0)}(0)=\frac{2\,  \Pi^{-1}-1}{\pi}.
 \label{h0}
\end{equation}
Indeed, imposing this condition to the Lorentzian and  Gaussian distributions, 
we recover (\ref{g_infty}) and (\ref{nu_infty}), respectively.
As expected, the region of stable incoherence widens as $\Pi$ decreases,
since in the limit $\Pi\to0$ the contribution of the first harmonic vanishes.
Equation \eref{h0} is a generalization for arbitrary
$\Pi$ of $h_\infty(0)=\pi^{-1}$ for $\Pi=1$ \cite{MP11}.

\subsubsection{Off-centered $h(q)$ ($q_0\ne0$)} If the distribution of $q$ is not 
centered at zero, criterion \eref{h0} is not valid. 
Apart of solving equations \eref{primera} and \eref{segunda} numerically,
one may resort to perturbation theory for small values of $|q_0|$. 
To avoid further complications we adopt $\Pi=1$ in the calculation that follows ---we can rescale
\eref{primera} and \eref{segunda} by $2\Pi^{-1}-1$, and recover this factor at the end of the calculation.
Thus, let us define first an even function $\tilde h$  setting the origin at $q_0$,
\begin{equation}
\tilde h(q)=h(q+q_0) .
\end{equation}
Equations \eref{primera} and \eref{segunda} become then:
\numparts
\begin{eqnarray}
\frac{1}{\pi (1+\Lambda^2)}= \tilde h_\infty(-q_0-\Lambda), \label{primeras}\\
\Lambda=-(1+\Lambda^2) \int_{-\infty}^\infty \frac{\tilde h_\infty(q-q_0-\Lambda)}{q}  \, \rmd q .
\label{segundas}
\end{eqnarray}
\endnumparts
At criticality we expect a generalization of \eref{h0} of the form
\begin{equation}
\tilde h_\infty^{(q_0)}(0)=\tilde h_\infty^{(q_0=0)}(0)+ \eta(q_0),
\end{equation}
where $\tilde h_\infty^{(q_0=0)}(0)=\pi^{-1}$, and $\eta$ is an even function with $\eta(0)=0$.

Assuming small $|q_0|$ and $|\Lambda|$  and twice differentiability of  $\tilde h(q)$ we approximate
\eref{primeras} and \eref{segundas} at leading order 
\numparts
\begin{eqnarray}
0= \Lambda^2 +\eta(q_0) + \frac{\pi}{2}(\Lambda+q_0)^2  \frac{\rmd^2 }{\rmd q^2} \tilde h_\infty^{(q_0=0)}(0)  ,\label{primeras2}\\
\Lambda=(\Lambda+q_0) \, I ,
\label{segundas2}
\end{eqnarray}
\endnumparts
% \begin{align}
% 0&= \Lambda^2 +\eta(q_0) + \frac{\pi}{2}(\Lambda+q_0)^2  \frac{\rmd^2 }{\rmd
% q^2} \tilde h_\infty^{(q_0=0)}(0),  \label{primeras2}\\
% \Lambda&=(\Lambda+q_0) \, I,
% \label{segundas2}
% \end{align}
where  $I=\int_{-\infty}^\infty q^{-1} \frac{\rmd\tilde  h_\infty^{(q_0=0)}(q)}{\rmd q}  \, \rmd q$.
Then, after some algebra we get $\eta(q_0)=b q_0^2$, with the constant $b$:
\begin{equation}
 b=-\frac{I^2+\frac{\pi}{2}\frac{\rmd^2 }{\rmd q^2} \tilde
 h_\infty^{(q_0=0)}(0)}{(1-I)^2}.
\end{equation}
For the Lorentzian distribution $b=0$, in consistency with the independence of $\gamma_\infty$ on $q_0$. For the Gaussian
distribution $b=-(4-\pi)(2+\pi)^{-2}=-0.0325\dots$. In terms of $\nu_\infty$, and recovering the $(2\Pi^{-1}-1)$ factor,  this means:
\begin{equation}
\nu_\infty^{(q_0)}\simeq\frac{\Pi}{2-\Pi} \sqrt{\frac{\pi}{2}}
\left(1-\sqrt{\pi} b q_0^2 \right).
\end{equation}
This is the perturbative extension at order $q_0^2$ of \eref{nu_infty}, which
implies 
that unstable incoherence may
achieve larger values of $\nu$, i.e.~broader distributions.

%%%%%%%%%%%%%%%%%%%%%%%%%%%%%
\section{Exact analysis: Ott-Antonsen ansatz}
\label{sec:oa}
%%%%%%%%%%%%%%%%%%%%%%%%%%%%%

Our aim is now the study of the full Winfree model defined by (\ref{eq1})-(\ref{p}), with no other
approximation than the thermodynamic limit.
However, due to mathematically tractability we restrict our analysis to Lorentzian heterogeneities (\ref{lorentzians}). 
The stability boundary of asynchrony in the $(\Delta,\varepsilon)$ plane is obtained below for different values 
of $q_0$, $\gamma$, and $r$.
An interesting question is to elucidate 
how the critical value of PRC 
heterogeneity $\gamma_\infty$ found in the averaged model 
translates into the full model.
Recalling  that  $\Pi(r)=(1+r)/2$ for the pulse shape \eref{p}, \eref{g_infty} yields:
\begin{equation}
\gamma_\infty=\frac{1+r}{3-r}. 
\label{g_infty_r}
\end{equation}
The averaged model in the preceding section predicts that for $\gamma>\gamma_\infty$, asynchrony is always stable, and the full model
must agree with this in the weak coupling limit. We anticipate that the results that follow are
perfectly consistent with \eref{g_infty_r}, but the model will achieve this consistency
in a different way depending on the sign of $q_0$.

\subsection{Derivation of low-dimensional equations}

As in section \ref{sec:av}, we adopt the thermodynamic limit $N\to\infty$
and define the density function $F(\theta|\omega,q,t)$.
This function  obeys the continuity equation:
\begin{equation}
 \partial_t F =- \partial_\theta\left\{\left[ \omega+\varepsilon Q(\theta) H(t)
 \right]F\right\},
\label{cont}
\end{equation}
where $H(t)$ is the mean field
\begin{equation}
% h(t)=\iint_{-\infty}^\infty p(\omega,q)  \int_0^{2\pi} F(\theta|\omega,q,t) \, P(\theta) \, d\theta \, d\omega \, dq
H(t)=%\iint_{-\infty}^\infty p(\omega,q)   
\left<
\int_0^{2\pi} F(\theta|\omega,q,t) \, P(\theta) \, d\theta 
\right>.
\label{H}
\end{equation}

For the theoretical analysis that follows 
we assume that $F$  satisfies the Ott-Antonsen ansatz \cite{OA08}:
\begin{equation}
F(\theta|\omega,q,t)=\frac{1}{2\pi}
\left\{ 1+ \left[\sum_{m=1}^{\infty} \alpha(\omega,q,t)^m \rme^{\rmi m\theta}  +
\mbox{c.c.}\right] \right\}.
\label{oa}
\end{equation}
Here, $\alpha^*$ is the first Fourier mode of the density, and therefore:
\begin{equation}
  Z^*(t)=\left< \alpha(\omega,q,t)\right>.
\label{z*}  
\end{equation}
The Ott-Antonsen ansatz can be applied 
to the Winfree model (\ref{eq1}),
with the PRC distributed according to (\ref{prc}),
since the model belongs to the family of phase models that can be written in
the form: 
\begin{equation}
\dot\theta(\mathbf{x},t)= B(\mathbf{x},t) + \mathrm{Im}\left[G(\mathbf{x},t) \rme^{-\rmi\theta(t)}\right] , 
\end{equation}
where $\mathbf{x}$ is a vector containing different parameters
that are distributed \cite{PR11,PM14,PD16}. 
In our case $\mathbf{x}=(\omega,q)$, with
$B(\mathbf{x},t)=\omega+\varepsilon q H(t)$ and
$G(\mathbf{x},t)= \varepsilon (1-\rmi q) H(t)$.
It has been shown that, 
if $F$ does not initially satisfy (\ref{oa}), it subsequently converges to it
---in the sense of \cite{OA09,OHA11}. 
Theoretical studies \cite{vlasov16} 
suggest that finiteness of the population cannot be expected to 
drive the system away from the OA manifold, 
and hence the formulation in terms of densities is reliable. 
Since the original article of Ott and Antonsen 
this has been confirmed numerically in a large number works, see e.g.~\cite{PD16} and references therein. 

Inserting \eref{oa} into the continuity equation (\ref{cont}) we get an equation for $\alpha(\omega,q,t)$:
\begin{equation}
\partial_t \alpha  = -\rmi \omega \alpha
+ \frac{\varepsilon H}{2} \left[1-\alpha^2 + \rmi q (1-\alpha)^2 \right].
% \partial_t \alpha  = -\rmi(\omega+ \varepsilon h q) \alpha
% + \frac{\varepsilon H}{2} \left[(1+\rmi q) -(1-\rmi q)\alpha^2 \right] 
\label{alpha} 
\end{equation}
Note that every $\alpha(\omega,q,t)$ is coupled with all others $\alpha(\omega',q',t)$
through the mean field $H$, see \eref{H}.
It was found in \cite{Gallego2017} ---see also the Supplemental Material of \cite{MP18}---
that for the pulse type \eref{p} $H$ is related
with $Z$ via
\begin{equation}
H(Z)=\mathrm{Re} \left[ \frac{1+Z}{1-rZ} \right].
\label{hz}
\end{equation}
To proceed further with the analysis, we note that the equation governing $|\alpha|$ is
\begin{equation}
\partial_t |\alpha| = \frac{\varepsilon H}{2} (\cos\phi+q\sin\phi)
\left(1-|\alpha|^2\right),
\label{abs_alpha}
\end{equation}
where $\phi=\arg(\alpha)$.
As the velocity vanishes at $|\alpha|=1$, $\alpha$ cannot leave the unit disk  ---otherwise \eref{oa} is not convergent. 
In close analogy to previous work \cite{OA08,MP11}  
the next key observation 
is that $\alpha$ admits an analytic continuation into 
the lower half complex $\omega$-plane, and the lower half complex $q$-plane (for positive $\varepsilon$).
If the field $\alpha(\omega,q,t)$ admits an analytic continuation at $t=0$, 
this will be the case for $t>0$ since $\alpha$ obeys the differential equation \eref{alpha} \cite{CL}. 
The  complexification of $\omega=|\omega|\rme^{\rmi\xi}$ and $q=|q| \rme^{\rmi \vartheta}$, transforms
\eref{abs_alpha} into: 
 \begin{eqnarray}
\fl \partial_t |\alpha| =|\omega| |\alpha| \sin\xi + \frac{\varepsilon H}{2} \left\{\cos\phi \left(1-|\alpha|^2\right) \right. \nonumber \\
\left.+ |q|\left[ \sin(\phi-\vartheta)+2|\alpha|\sin\vartheta
 -|\alpha|^2 \sin(\phi+\vartheta) \right] \right\}.
 \end{eqnarray}
At $|\alpha|=1$ the velocity is
\begin{equation}
\partial_t |\alpha| =|\omega|  \sin\xi+ \varepsilon H |q| \sin\vartheta
(1-\cos\phi).
\end{equation}
Provided $\sin\xi\le0$, and $\sin\vartheta\le0$ (for positive $\varepsilon$),
$\partial_t |\alpha|\le0$, and therefore $\alpha$ cannot leave the unit disk, if initially inside.

The analytic continuation of $\alpha$ allows
to apply twice the residue's theorem to the integrals in \eref{z*} by closing the respective
integration contours by large semicircles in the lower half $\omega$- and $q$-planes.
As the Lorentzian distribution
has only one pole inside the integration contour,
a simple relation between $Z$ and
$\alpha$ is found:
\begin{equation}
  Z^*(t)=\alpha(\omega_p,q_p,t),
\end{equation}
where $\omega_p=\omega_0-\rmi\Delta$ and $q_p=q_0-\rmi\gamma$ are the poles of $g(\omega)$ and $h(q)$, respectively.
Hence we only have to evaluate \eref{alpha} at $(\omega_p,q_p)$, in order to obtain
one complex-valued ODE for $Z$:
\begin{equation}
% \dot Z  = i(\omega_p^*+ \varepsilon h q_p^*) Z
% + \frac{\varepsilon h}{2} \left[(1-iq_p^*) -(1+iq_p^*) Z^2 \right] 
\dot Z  = \rmi \omega_p^* \, Z
+ \frac{\varepsilon H(Z)}{2} \left[1-Z^2 - \rmi q_p^* \, (1-Z)^2 \right],
 \label{Z} 
 \end{equation}
where $H(Z)$ is given by \eref{hz}.
\Eref{Z} completely describes the asymptotic dynamics of the model (in the thermodynamic limit).
Hereafter, we set $\omega_0=1$, since this can be achieved through
trivial rescalings of time, $\Delta$ and $\varepsilon$ by $\omega_0>0$ in \eref{Z}.

%%%%%%%%%%%%%%%%%%%%%%%%%%%%%
\subsection{Analysis of the low-dimensional system \eref{Z}}
%%%%%%%%%%%%%%%%%%%%%%%%%%%%%

\Eref{Z} is a planar system, generically with two possible attractor types: fixed point and
limit cycle. Our previous work with homogeneous PRCs \cite{PM14,Gallego2017} revealed that 
the model may exhibit two simultaneously stable fixed points, and that limit cycles correspond to partially
synchronized states. For small coupling, in particular, only one fixed point with $|Z|\ll1$  (asynchrony)
exists, which corresponds to the incoherent solution $Z=0$ of the averaging approximation \eref{av_winfree2}. 
We focus next on the stability boundary of the asynchronous state, 
which is determined applying the {\sc matcont} toolbox of {\sc matlab} to \eref{Z}.

\subsubsection{Dirac delta pulses.}

\begin{figure}
  \includegraphics*[width=0.99\textwidth]{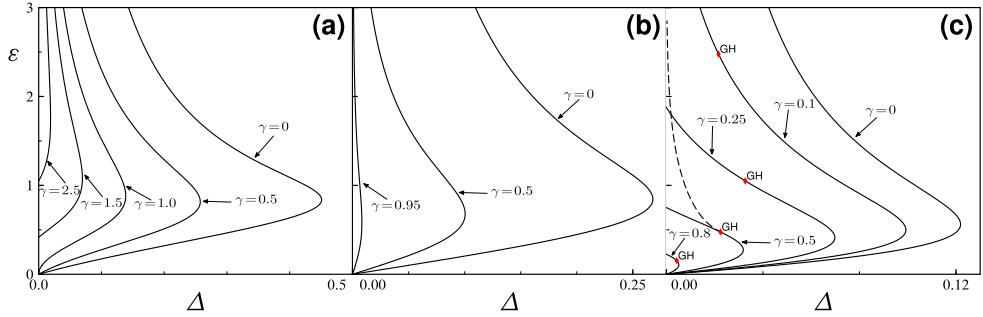}  
  \caption{Stability boundary of asynchrony when the distribution of PRCs is centered at (a) $q_0=1$, (b) $q_0=0$, and (c) $q_0=-1$. 
Asynchrony is unstable at the left of the solid lines. The pulse is $P(\theta)=2\pi\delta(\theta)$.}
  \label{fig::dirac}
\end{figure}

As reference case, let us determine first the stability boundary of asynchrony 
for the Dirac delta pulse, $r=\Pi=1$, and  in
the absence of PRC diversity, $\gamma=0$. As depicted in figure \ref{fig::dirac} for $q_0=1,0,-1$, the stability boundary of
asynchrony is a line in the $(\Delta,\varepsilon)$ plane that emanates 
from $(\Delta,\varepsilon)=(0,0)$ with a slope equal to $2$, 
as correctly predicted by the averaging approximation, see (\ref{av_c}).
This line is the locus of a (supercritical) Hopf bifurcation of asynchrony.
Contrary to what could be naively inferred from (\ref{av_c}),
the boundary is not a straight line: it folds back at a certain $\Delta$ value and approaches the $\varepsilon$-axis asymptotically as $\varepsilon\to\infty$.
This behavior is common to all $q_0$ values, see figure 9 in \cite{Gallego2017}.

Introducing heterogeneity in the PRCs must have an important effect, because ---according to the averaging
approximation--- incoherence is always stable for $\gamma>\gamma_\infty=1$. 
Strictly speaking,  this only applies
to small $\varepsilon$ and $\Delta$, where the averaging approximation is valid. 
As can be see in figure~\ref{fig::dirac}(a), for $q_0=1$, the instability boundary detaches from the origin when $\gamma$ exceeds $\gamma_\infty=1$. However,
as shown in figure~\ref{fig::dirac}(c), for $q_0=-1$ the disappearance of the boundary from the neighborhood of the origin occurs
in a completely different way: The domain of unstable asynchrony progressively shrinks as $\gamma$ grows, 
collapsing with the origin exactly when $\gamma=\gamma_\infty=1$.
We notice also that, in the $q_0=-1$ case, as $\gamma$ grows from zero a generalized Hopf (GH) point appear, in such a way that the Hopf boundary
is of subcritical type above that point. For $\gamma=0.5$ we depict with dashed line the locus of the saddle-node
bifurcation of limit cycles emanating from GH ---as for other $\gamma$ values, we skip this information.
Finally, for the singular case $q_0=0$, see figure \ref{fig::dirac}(b), the domain of unstable asynchrony shrinks
as $\gamma$ approaches $\gamma_\infty=1$, collapsing
with the entire $\varepsilon$-axis. Indeed for $q_0=0$ the 
exact boundary can be obtained in parametric form, but the formulas are convoluted and we skip them here.

Apart from the results in figure~\ref{fig::dirac} for particular $q_0$ values, the analytical
study of \eref{Z} permits to corroborate that the scenarios for $q_0=1$ and $q_0=-1$ apply, respectively, to
all positive and negative values of $q_0$.
For the analysis of \eref{Z}, we found it convenient to define a new complex variable 
$w\equiv x+iy=(1+Z)/(1-Z)$. This is a conformal mapping from the unit disk $|Z|\le1$ onto
the right half plane $x\ge 0$.
The ODEs for the real and imaginary parts of $w$ are:
\begin{equation}
\eqalign{
\dot x = \frac{\Delta}{2}(1-x^2+y^2) -   \, x \, y + \varepsilon \, (x+\gamma)
H(x,y), \cr
\dot y = -\frac{1-x^2+y^2}{2} - \Delta \,x \,y + \varepsilon \, (y-q_0) H(x,y) .}
\label{cartesian}
 \end{equation}
For the Dirac delta pulse $H$ turns out to be very simple: $H(x,y)=x$.
Still the system \eref{cartesian} is too convoluted to 
find a closed expression of the Hopf boundary.
Useful information can be obtained nonetheless setting $\Delta=0$,
in order to find out at which point the Hopf line intersects the $\varepsilon$-axis. 
After getting the fixed point $(x_*,y_*)$, with coordinates 
\begin{equation}
x_*=\frac{\varepsilon q_0 + \sqrt{1+\varepsilon^2(1+q_0^2+\gamma^2)+\varepsilon^4\gamma^2}}{1+\varepsilon^2}
\end{equation}
and $y_*=\varepsilon(x_*+\gamma)$, trivial calculations yield the nontrivial $\varepsilon$-intercept of the Hopf line:
% \begin{equation}
% \varepsilon_H=\frac{\gamma^2-1}{2\gamma q_0} 
% %  \begin{cases}
% % \mbox{for } \gamma<1 &\mbox{if } q_0<0 \\
% % \mbox{for } \gamma>1 &\mbox{if } q_0>0 
% %  \end{cases},
% \left\{\begin{array}{lr}
%        \mathrm{for } \, \gamma>1 &  \mathrm{if } \, q_0>0, \\      
%        \mathrm{for } \,  \gamma<1 &  \mathrm{if } \, q_0<0.
%   \end{array}\right.     \label{eh}
% \end{equation}
\begin{equation}
\varepsilon_H^{(\Delta=0)}=\frac{\gamma^2-1}{2\gamma q_0} , 
\end{equation}
which is only valid for $\varepsilon_H^{(\Delta=0)}>0$, i.e.~ $\gamma>1$ if $q_0>0$ or $\gamma<1$ if $q_0<0$.
This formula is in fully agreement with the results in figure \ref{fig::dirac}, and gives support to the
general distinction between positive, negative, and vanishing $q_0$ cases.

\begin{figure}
  \includegraphics*[width=120mm]{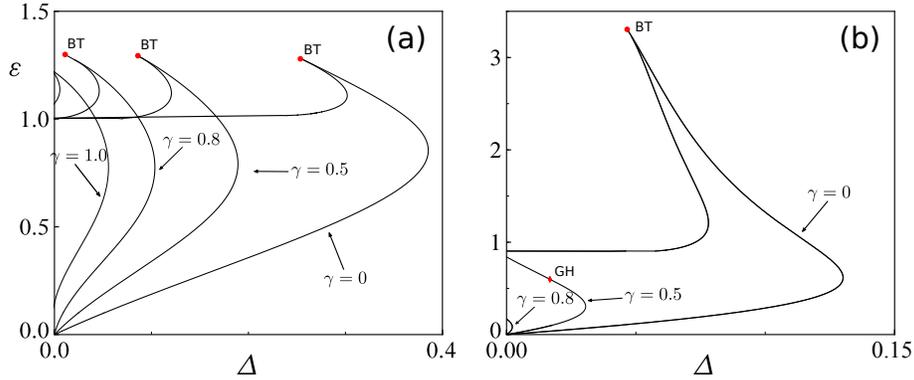}
  \caption{Stability boundary of the asynchronous state when the distribution of PRCs is centered at (a) $q_0=1$ and (b) $q_0=-1$. The pulse
  form is given by \eref{p} with $r=0.9$.}
  \label{fig::r09}
\end{figure}

\subsubsection{Pulse with finite width.}
When the pulse has finite width, in the absence of PRC diversity ($\gamma=0$), the asynchronous state is bounded by two
bifurcation lines: The supercritical Hopf-bifurcation line 
that emanates from the origin (with the slope predicted by the averaging approximation)
terminates at a double-zero eigenvalue, Bogdanov-Takens (BT), point, 
see e.g.~the lines for $r=0.9$ in panels (a) and (b) of figure \ref{fig::r09}. Additionally, from the BT point 
up to the $\varepsilon$ axis, a line corresponding to a saddle-node
bifurcation bounds the region of unstable asynchrony in its upper part.
We decided to limit our presentation to $r=0.9$, 
a value corresponding to a quite narrow pulse, see figure \ref{fig::qp}(b), since sharp pulses
are often observed in reality.
As can be seen in 
the two panels of figure \ref{fig::r09}, the displacement of the lines as $\gamma$ grows from zero 
is clearly reminiscent of what is observed for Dirac delta pulses, but now the detachment ($q_0=1$)
or collapse ($q_0=-1$) of the synchronization region occurs
for a smaller $\gamma$ value, which, according to (\ref{g_infty_r}), is
$\gamma_\infty=1.9/2.1=0.90476\ldots$

\section{Conclusions}
\label{sec:concl}
In this work we have carried out the first exact analysis of the Winfree model with heterogeneous PRCs. 
Analytical results for networks of coupled 
oscillators with heterogeneous PRCs are scarce. Even though the model investigated here bears a 
strong resemblance with that of \cite{tsubo07}, the two works are 
hardly comparable because of the different
parametrizations of the PRCs\footnote{In \cite{tsubo07} $Q_i(\theta)=\cos(a_i\pi)-\cos(\theta-a_i\pi)$, where $a_i$
is the distributed parameter. When this $Q_i(\theta)$ written in a form closer to \eref{prc},
$Q_i(\theta)=\cos(a_i\pi)(1-\cos\theta)-\sin(a_i\pi)\sin\theta$, it becomes evident that no direct mapping
between $q$ and $a$ distributions exists.} and the discontinuous coupling function used there. 

In the first part of our paper, we showed that 
the averaging approximation of the Winfree model with heterogeneous PRCs and Dirac delta pulses ($\Pi=1$) turns out to 
be the Kuramoto model with distributed shear~\cite{MP11}. 
We found that, under the averaging approximation, 
the incoherent state becomes always stable beyond a critical level of PRC heterogeneity.
These results hold for general distributions of heterogeneity, and different pulse widths ($\Pi$ values).
  
In the second part we analyzed the full model. To 
achieve the maximal dimensionality reduction with the Ott-Antonsen ansatz
we restricted our analysis to Lorentzian distributions.  
The system of two ODEs obtained describes the system exactly in the thermodynamic limit.
We found that 
the sign of parameter $q_0$, controlling the offset of the
PRC distribution, plays a fundamental role in the response of the system against 
PRC heterogeneity.

In future work, nonindependent joint distributions of $\omega$ and $q$
could be explored following \cite{PM11}. 
Adaptation-mediated changes in the PRCs appears to be another plausible line of research.
In contrast, changing the mean-field interactions 
by short-range, long-range or networked interactions is quite a challenge.

\ack
We acknowledge support by MINECO (Spain) under Projects 
No.~FIS2016-74957-P, No.~PSI2016-75688-P 
and No.~PCIN-2015-127. 
We also acknowledge support 
by the European Union's Horizon 2020 research and innovation
programme under the Marie Sk{\l}odowska-Curie grant agreement No.~642563.

% \bibliography{bibliografia}
% \bibliographystyle{jphysicsB}

\end{document}